\documentclass[
    ,final            
  ]
  {aipproc}

\usepackage{amssymb}
\usepackage{bbm}
\usepackage{fancyheadings}
\usepackage{epsfig}
\usepackage{graphicx}     
\usepackage{dcolumn}      
\usepackage{psfig,pstricks,pst-grad,fancybox,graphics}
\usepackage{enumerate}
\usepackage[latin1]{inputenc} 

\newcommand{\R}{\ensuremath{\mathbbm R}}

\newcommand{\bra}[1]{\ensuremath{\langle#1|}}

\newcommand{\ket}[1]{\ensuremath{|#1\rangle}}

\newcommand{\braket}[2]{\ensuremath{\langle #1|#2\rangle}}
\newcommand{\KetBra}[1]{\ensuremath{| #1 \rangle \langle #1 |}}
\newcommand{\ketbra}[1]{\ensuremath{| #1 \rangle \langle #1 |}}

\newcommand{\Eins}{\ensuremath{\mathbbm 1}}
\newcommand{\eins}{\ensuremath{\mathbbm 1}}

\newcommand{\WW}{\ensuremath{\mathcal{W}}}
\newcommand{\PP}{\ensuremath{\mathcal{P}}}

\newcommand{\BE}{\begin{equation}}
\newcommand{\EE}{\end{equation}}
\newcommand{\be}{\begin{equation}}
\newcommand{\ee}{\end{equation}}
\newcommand{\bea}{\begin{eqnarray}}
\newcommand{\eea}{\end{eqnarray}}
\newcommand{\kommentar}[1]{}

\newcommand{\mean}[1]{\ensuremath{\langle #1 \rangle}}

\newcommand{\vr}{\ensuremath{\varrho}}

\layoutstyle{8x11single}

\begin{document}

\title{Separability Criteria from Uncertainty Relations}

\author{Otfried G\"uhne}{
  address={Institut f\"ur Quantenoptik und Quanteninformation, 
\"Osterreichische Akademie der Wissenschaften, A-6020~Innsbruck, Austria
}
,altaddress={Institut f\"ur Theoretische Physik, Universit\"at Hannover,
Appelstra{\ss}e 2, D-30167 Hannover, Germany}
}

\author{Maciej Lewenstein}{
address={Institut f\"ur Theoretische Physik, Universit\"at Hannover,
Appelstra{\ss}e 2, D-30167 Hannover, Germany}
}

\begin{abstract}
We explain several separability criteria which rely on 
uncertainty relations. For the derivation of these criteria
uncertainty relations in terms of variances or entropies 
can be used. We investigate the strength of the 
separability conditions for the case of two qubits and 
show how they can improve entanglement witnesses.
\end{abstract}

\maketitle


\section{Introduction}
Entanglement is one of the counterintuitive phenomena of 
quantum mechanics and is, despite a lot of progress in the 
last years, not fully understood. A state 
$\vr$ on a bipartite system is called {\it separable} when it
can be written as a convex combination of product states, 
i.e., $ \vr=\sum_i p_i \ketbra{a_i b_i},$ where $p_i \geq 0$ 
and $\sum_i p_i=1.$
A state which is not separable is called {\it entangled.}
The question, whether a state $\vr$ is entangled or not, 
is the so-called {\it separability problem,} and no general 
answer to this question is known (for a review see 
\cite{alberbuch}). Geometrically, the definition 
of separability implies that the separable states 
form a {\it convex set} in the (high dimensional)
real vector space of all density matrices of a 
given system (see Fig.~1(a)).

Given a state $\vr,$ what shall we measure 
to detect the entanglement in this state? This is a 
question of great importance in many experiments, since 
the presence of entanglement is a necessary 
precondition for certain tasks as quantum key 
distribution or teleportation. The usual tools 
to answer this question are
{\it entanglement witnesses} (EW) \cite{horoppt}. An EW $\WW$ is 
a Hermitean observable $\WW$ with an positive expectation 
value on all separable states. Thus $Tr(\WW\varrho)<0$
implies that the state $\vr$ is entangled.  
An EW provides a criterion, which depends linearly 
on the state. Geometrically, the set where $Tr(\WW\varrho)=0$ 
is a hyperplane, separating the detected states 
from the non-detected ones.

Can we use {\it nonlinear} witnesses?
This is a very natural question for the following reason: 
One might expect that one can approximate the convex set of 
the separable states better by using a nonlinear expression
(see Fig.~1(a)). 
This paper deals with special types of nonlinear witnesses, 
and in fact, it will turn out that sometimes they can improve 
already known linear witnesses. The nonlinear expressions we will
use are based on uncertainty relations. 

\section{Uncertainty based criteria}
Let us start with criteria based on variances of observables.
The variance of an observable $M$ in the state $\vr$ is given by 
$
\delta^2(M)_{\varrho} := \mean{(M-\mean{M}_{\varrho})^2}_{\varrho}
= \mean{M^2}_{\varrho}-\mean{M}^2_{\varrho},
$
where $\mean{M}_\vr=Tr(\vr M).$
If $\vr=\ketbra{\psi}$ describes a pure state, the variance of $M$
is zero iff $\ket{\psi}$ is an eigenstate of $M.$ Furthermore,
the variance is concave in the state. If 
$\varrho=\sum_k p_k \varrho_k$ is a convex combination of some 
states $\varrho_k,$ then
\begin{equation}
\sum_i \delta^2(M_i)_{\varrho} 
\geq 
\sum_k p_k \sum_i \delta^2(M_i)_{\varrho_k}
\label{l1a}
\end{equation}  
holds. This inequality can be straightforwardly calculated 
\cite{hofmann1, guehne1}, and expresses the simple physical 
fact that one cannot decrease the uncertainty of an observable 
by mixing several states. 
H.~Hofmann and S.~Takeuchi were the first who realized that
this property of the variance gives rise to separability 
criteria, the so-called {\it local uncertainty relations} (LURs). 
They showed the following:

{\bf Criterion 1} \cite{hofmann1}. Let $A_i,B_i,  i=1,...,n$  be 
operators on Alice's (respectively,  Bob's) space, fulfilling 
\be
\sum_{i=1}^n \delta^2(A_i)\geq U_A \;\;\;\mbox{    and    } \;\;\;
\sum_{i=1}^n \delta^2(B_i)\geq U_B.
\label{lur1}
\ee
We define $M_i := A_i \otimes \Eins +\Eins \otimes B_i.$ 
Then we have for a separable $\vr$ the inequality
\begin{equation}
\sum_{i=1}^n\delta^2(M_i)_{\varrho} \geq U_A + U_B.
\label{lur2}
\end{equation}
These criteria have a beautiful and clear physical interpretation.
The Eqs.~(\ref{lur1}) are just uncertainty relations, expressing
the fact that the $A_i$ and $B_i$ do not share a common eigenstate. 
Then, Eq.~(\ref{lur2}) shows that the separable states 
inherit the bounds from the local uncertainty relations in 
Eqs.~(\ref{lur1}).
In Ref.~\cite{guehne1} these criteria were generalized in the 
following way:

{\bf Criterion 2} \cite{guehne1}. A state $\vr$ is entangled if there 
exist $M_i$ and a constant $C>0$ such that
$\sum_i \delta^2(M_i)_\varrho < C,$ holds,
while for all product states
$\sum_i \delta^2(M_i)_{\varrho_s} > C$
is valid.

Although this criterion looks fairly obvious, it turnes out that
a proper choice of the $M_i$ guarantees to detect many entangled 
states. For instance, all pure bipartite entangled states and a 
family of bound entangled states can be detected \cite{guehne1}. 
Also, one can detect multipartite entanglement and relate
the variance based criteria for finite dimensional systems with 
criteria for infinite dimensional systems. But these connections 
are beyond the scope of this paper.

\begin{figure}
\setlength{\unitlength}{1cm}
\begin{picture}(15.5,5.3)
\thinlines
\put(0,0){\includegraphics[height=5.3cm]{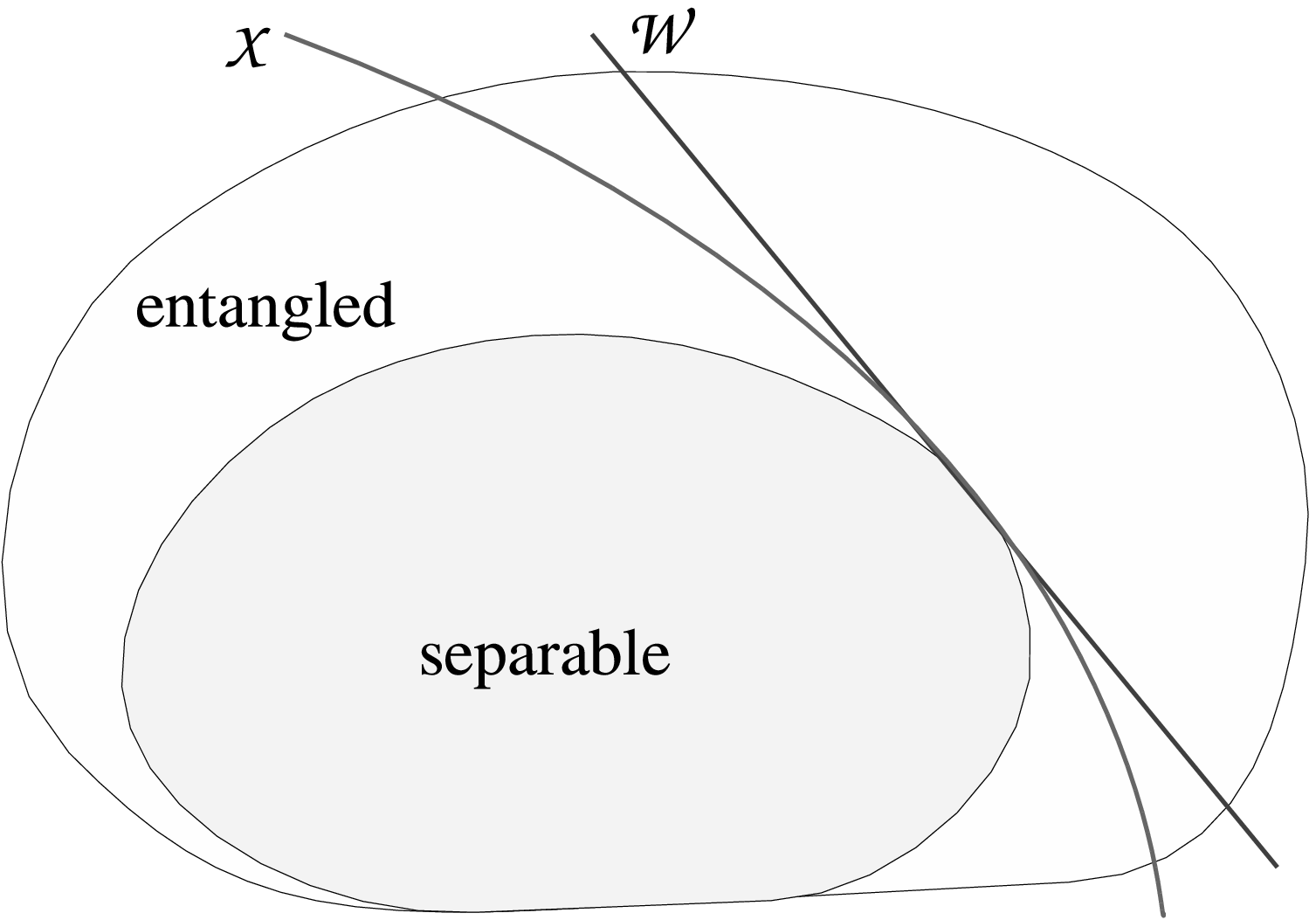}}
\put(8,-0.0){\includegraphics[height=5.3cm]{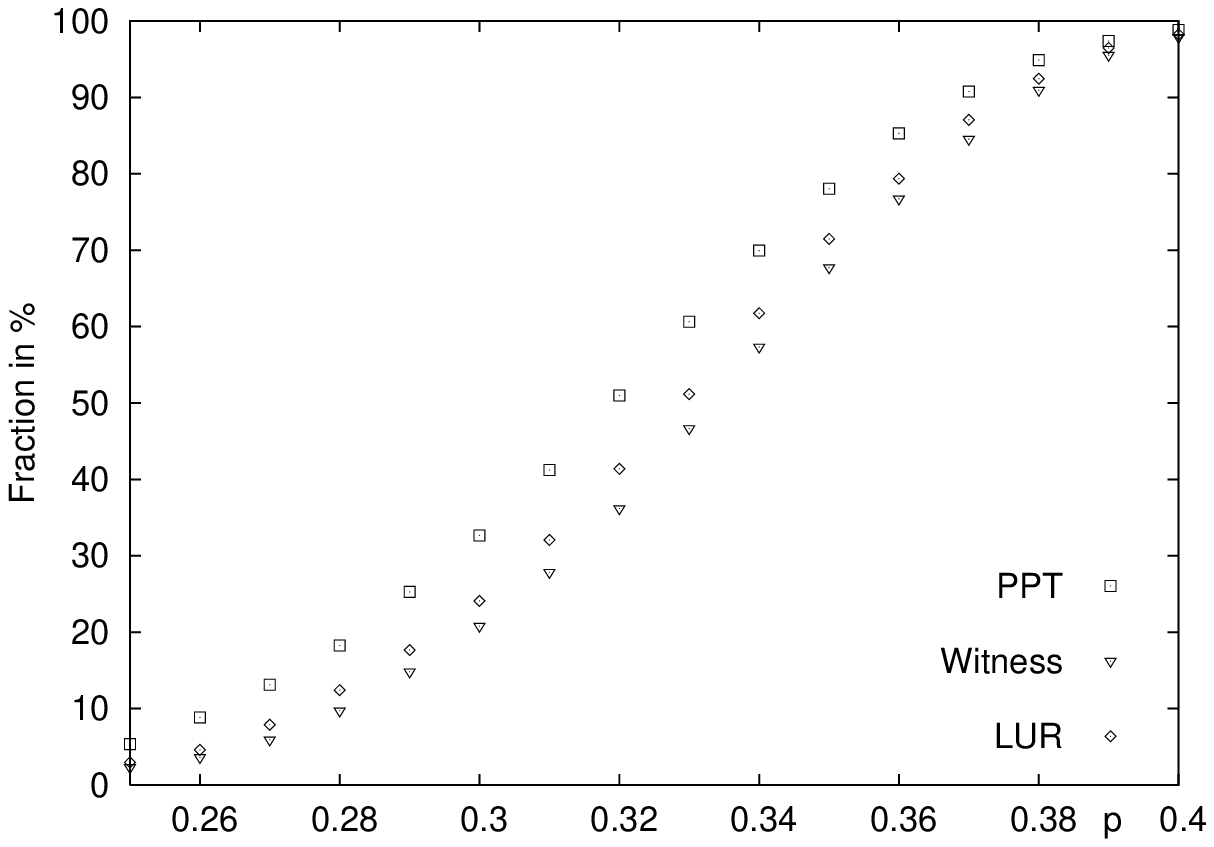}}
\put(0,4.8){\mbox{(a)}}
\put(7.8,4.8){\mbox{(b)}}
\end{picture}
\caption{ 
(a) Schematic view of the set of all states and the separable 
states as a convex subset. $\WW$ denotes a witness 
(i.e. the set where $Tr(\vr \WW)=0$) and $\mathcal{X}$ 
a possible nonlinear witness.
(b) A comparison between the LUR in Eq.~(\ref{lur5}), 
the witness resulting from the linear part of the LUR, and the 
PPT criterion for the family of states described in the text. 
In dependence on $p$ the fraction of states which
are detected via the different criteria is shown.
}
\end{figure}

A different way of using uncertainty relations to detect 
entanglement uses {\it entropic uncertainty relations} (EURs).
Let us briefly recall what these are. If we have a non-degenerate 
observable $M = \sum_i \mu_i \ketbra{m_i},$ a measurement of this 
observable in a quantum state $\varrho$ gives rise to a probability 
distribution of the different outcomes:
\begin{equation}
\PP(M)_\varrho=(p_1, ..., p_n); \;\;\;
p_i=\bra{m_i}\varrho\ket{m_i}.
\label{maciej4}
\end{equation} 
It is now possible to measure the uncertainty of this measurement
by taking the entropy of this probability distribution, i.e., 
by defining $S(M):=S(\PP(M))_\varrho.$ The entropy $S$ used in 
this definition may be the standard {\it Shannon entropy}
$S^S(\PP):= - \sum_k p_k \ln(p_k),$
or, more generally any so-called {\it entropic function}
$S(\PP)=\sum_i s(p_i)$
where $s:[0;1] \rightarrow \R$ is a concave function, may be used.
An example is the {\it Tsallis entropy}
$S^T_q(\PP):=(1- \sum_k (p_k)^q)/(q-1),$ which depends on a parameter
$q>0,$ for $q=1$ we have $S^T_1=S^S.$ With this definition of the 
uncertainty of a measurement it is clear that for two observables 
$M = \sum_i \mu_i \ketbra{m_i}$ and  $N = \sum_i \nu_i \ketbra{n_i}$ 
which do not share a common eigenstate, there must exist a strictly 
positive constant $C$ such that
\begin{equation}
S^S(M)+S^S(N) \geq C
\label{eur}
\end{equation}
holds. Estimating $C$ is not easy, however it was shown 
in Ref. \cite{maassen} that one could take 
$C= - 2 \ln (\max_{i,j} |\braket{m_i}{n_j}|).$
When $M$ is a degenerate observable the definition of 
$S(M)$ in the previous way is not applicable, since
the spectral decomposition is not unique in this case.
However, there is a unique way in writing $M=\sum_i \mu_i X_i$
where the $\mu_i$ are pairwise different and where there 
$X_i$ are now projectors onto the corresponding eigenspaces.
Then on can define $\PP$ by $p_i=Tr(\varrho X_i)$ and 
finally $S(M).$

The entropy $S(M)$ fulfills a similar concavity property as 
the variance in Eq.~(\ref{l1a}). This can be used to detect 
entanglement in a similar manner, as it was recognized in 
Ref.~\cite{giovannetti}. Later, in Ref.~\cite{guehne2} the 
following separability criteria in terms of EUR were shown:

{\bf Criterion 3} \cite{guehne2}. Let $A_1,A_2,B_1,B_2$ be 
observables with nonzero eigenvalues on Alice's (respectively,  
Bob's) space obeying an EUR of the type 
\begin{equation}
S(A_1)+S(A_2)\geq C
\end{equation}  
or the same bound for $B_1,B_2.$ If $\varrho$ is 
separable, then 
\begin{equation}
S(A_1 \otimes B_1)_\varrho + 
S(A_2 \otimes B_2)_\varrho 
\geq C.
\end{equation}

For entangled states this bound can be violated, since 
$A_1\otimes B_1$ and $A_2\otimes B_2$ might be degenerate 
and have a common entangled eigenstate. This criterion
shows how any EUR on one part of the system results in a 
separability criterion on the composite system.  
This property is similar to the construction of the LURs. 
For a detailed investigation of this criterion we refer 
to \cite{guehne2}. A different 
criterion can be established when only one observable, 
but with entangled eigenstates, is considered. 
Then, the entropy of its
measurement cannot vanish for separable states.  
So one 
can prove:

{\bf Criterion 4} \cite{guehne2}. Let 
$M=\sum \mu_i \ketbra{m_i}$ be a non degenerate observable. 
Let $c<1$ be an upper bound for all the squared Schmidt 
coefficients of all $\ket{m_i}.$ Then for all separable states
\begin{equation}
S^T_q(M) \geq 
\frac{1- \lfloor 1/c \rfloor c^q - (1-\lfloor 1/c \rfloor c)^q}{q-1}.  
\label{stbound}
\end{equation}

\section{Investigation of the  criteria}

Let us start with an investigation of the LURs. For a single-qubit 
system is has been shown \cite{hofmann1} that for the Pauli matrices 
the uncertainty relation
$ \sum_{i=x,y,z} \delta^2(\sigma_i) \geq 2$
holds. Defining $M_i=\sigma_i \otimes \eins + \eins\otimes \sigma_i$
this yields the LUR $\sum_{i=x,y,z} \delta^2(M_i) \geq 4.$
A short calculation shows that from  this equation it follows
that for all separable states
\begin{equation}
\mean{ 
\eins \otimes \eins + 
\sigma_x \otimes \sigma_x+
\sigma_y \otimes \sigma_y+
\sigma_z \otimes \sigma_z
} 
-\frac{1}{2} 
\sum_{i=x,y,z} \mean{\sigma_i \otimes \eins + \eins\otimes \sigma_i}^2 
\geq 0
\label{lur5}
\end{equation}
has to hold. This is a quite remarkable equation for the following 
reason. The first part, which is linear in the expectation values 
is known to be an {\it optimal} entanglement witness \cite{guehnepra}. 
{From} this witness some quadratic terms are subtracted. Thus, 
in this case, the LUR can be viewed as a nonlinear EW which improves 
a linear EW. 

Let us investigate how big the improvement is. To this
aim, we look at states of the form
$
\varrho(p,d):= p \KetBra{\psi^-} + (1-p)\sigma, 
$
where 
$
\Vert \sigma - \frac{1}{4}\Eins \Vert \leq d
$ 
and $\Vert A \Vert = \sqrt{Tr(A A^\dagger)}.$
Physically, these states are a mixture of a singlet 
state and some separable noise, the parameter $p$ determines 
the fidelity of the singlet state, and the parameter $d$ the 
properties of the noise. For $d=0$ the noise consists of white 
noise. The set of matrices $\varrho(p,d)$ governs a ball in the 
space of all matrices. We take the value $d=0.2$ and generate 
matrices of the form $\varrho(p,0.2)$ randomly distributed in 
this ball \cite{karolz}. 
Then we investigate the separability properties of these
matrices. We determine the fraction of matrices which are 
detected by the witness and the LUR and check whether the
matrices have a positive partial transpose (PPT), which is 
necessary and sufficient for entanglement in this case 
\cite{horoppt, ppt}. Of course, these fractions depend on the value of 
$p.$ The results are shown in Fig.~1(b). On can clearly
see that the LUR improves the witness significantly, although 
it is not capable of detecting all states. 

\begin{figure}

\setlength{\unitlength}{1cm}
\begin{picture}(14,4)
\thinlines
\put(0,0){\includegraphics[height=4cm]{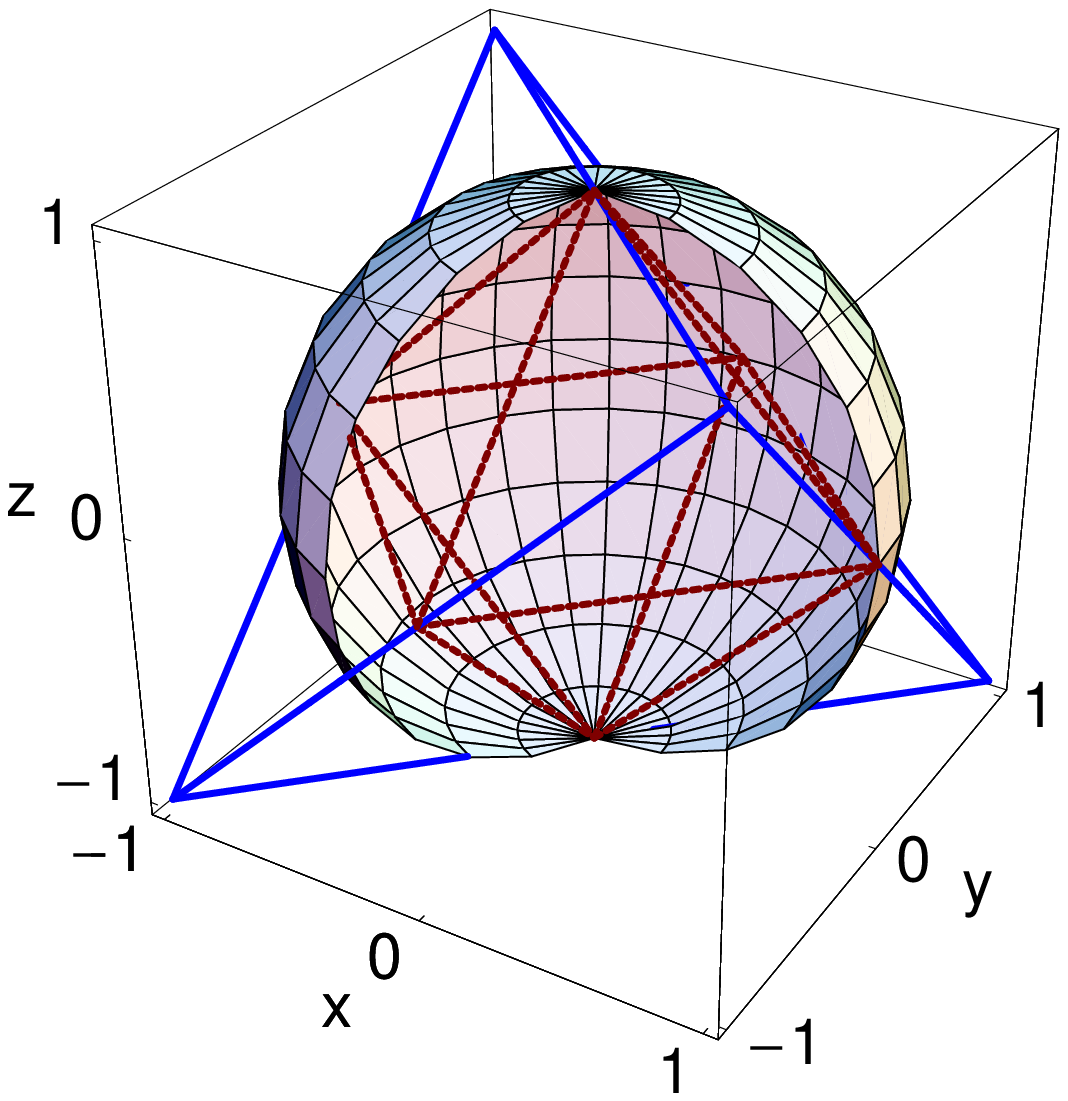}}
\put(5,0){\includegraphics[height=4cm]{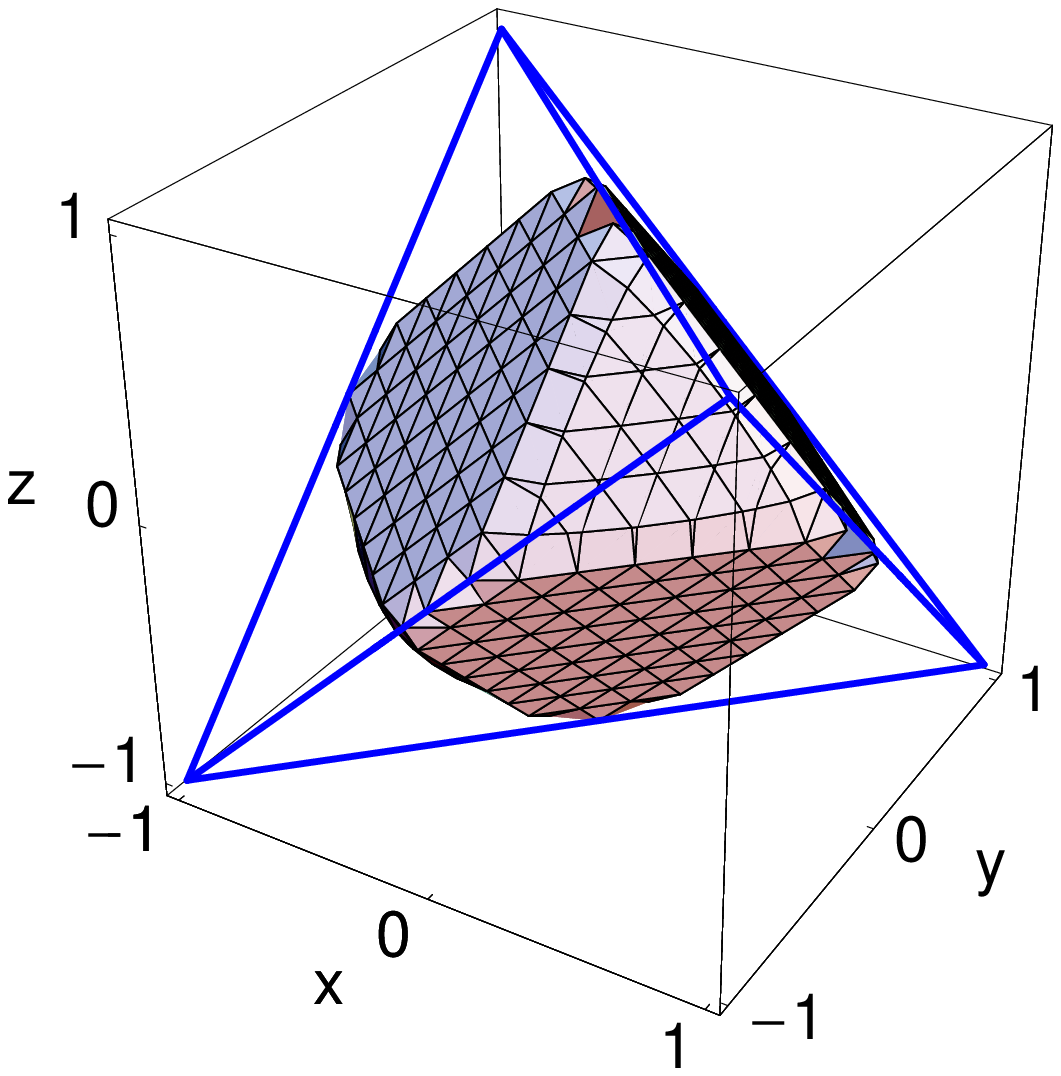}}
\put(10,0){\includegraphics[height=4cm]{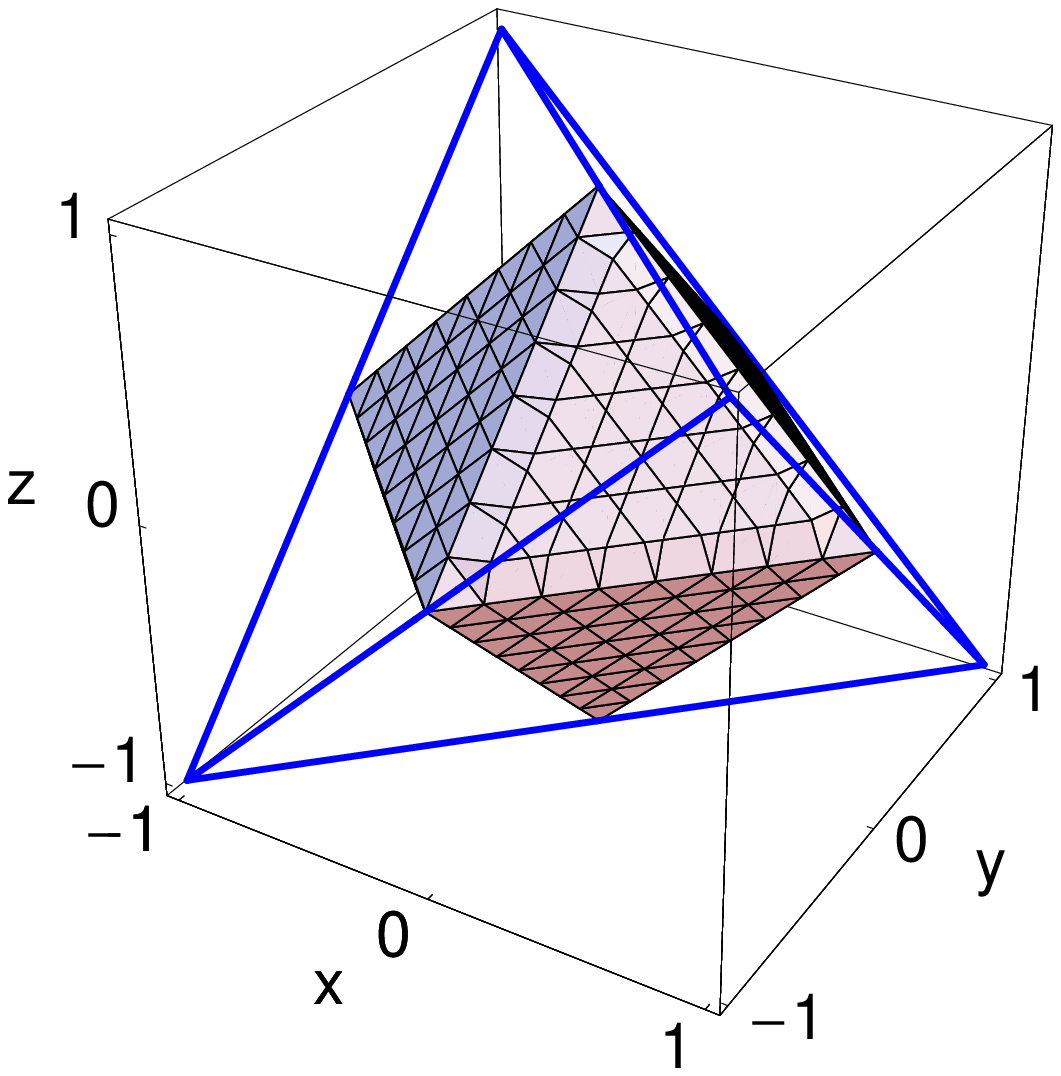}}
\put(0,0.5){\mbox{(a)}}
\put(5,0.5){\mbox{(b)}}
\put(10,0.5){\mbox{(c)}}
\end{picture}
\caption{ 
(a) The tetrahedron (solid lines) of all states and 
the octahedron (dashed lines) which contains the 
separable states. The ball represents the states which 
are not detected by Eq.~(\ref{gi1}).
(b) The subset of states which are not detected by Eq.~(\ref{zqb}) for $q=4.$
(c) As (b) but for $q=15.$ 
}
\end{figure}

To give a simple example for Criterion 2  we take as 
observables projectors onto Bell states. Let us
denote them by
$ 
\ket{BS_1}=(\ket{00}+\ket{11})/\sqrt{2}, 
\ket{BS_2}=(\ket{00}-\ket{11})/\sqrt{2}, 
\ket{BS_3}=(\ket{01}+\ket{10})/\sqrt{2},
\ket{BS_4}=(\ket{01}-\ket{10})/\sqrt{2}. 
$
{From} the fact that the Schmidt coefficient of the Bell states are 
$1/\sqrt2$ one can calculate that for separable states
\begin{equation}
\sum_i\delta^2(\ketbra{BS_i})_\vr = 
1 -\sum_i Tr(\vr \ketbra{BS_i})^2 
\geq \frac{1}{2}
\label{gi1}
\end{equation}
holds \cite{guehne1}. To investigate the strength and the geometrical 
meaning  of this inequality, let us introduce the coordinates 
$i=Tr(\varrho \sigma_i\otimes\sigma_i)$ for $i=x,y,z.$ In these 
coordinates we have
$
\bra{BS_1}\varrho \ket{BS_1}=(1+x-y+z)/4,
\bra{BS_2}\varrho \ket{BS_2}=(1-x+y+z)/4, 
\bra{BS_3}\varrho \ket{BS_3}=(1+x+y-z)/4,
$
and
$
\bra{BS_4}\varrho \ket{BS_4}=(1-x-y-z)/4.  
$

How does the state space look in these coordinates? 
An arbitrary density matrix has to obey 
$Tr(\varrho\ketbra{BS_i}) \geq 0,$ which leads to 
the four conditions $x-y+z \leq 1, -x+y+z \leq 1,
x+y-z \leq 1,$ and $-x-y-z \leq 1.$ These conditions describe a 
tetrahedron in the three-dimensional space (see Fig.~2(a)).
A separable state has to obey in addition 
$Tr(\varrho\WW^{(i)})\geq 0$ for the witnesses
$\WW^{(i)}=1/2 \cdot \eins-\ketbra{BS_i}.$ 
This results in $x+y+z \leq 1, -x-y+z \leq 1,
x-y-z \leq 1,$ and $-x+y-z \leq 1,$ which describes an 
octahedron in the tetrahedron from above. Furthermore, 
a straightforward calculation proves that in these coordinates 
Eq.~(\ref{gi1}) reads $ x^2+y^2+z^2 \leq 1.$
This is the equation of a three-dimensional sphere. 
The states inside this sphere are not detected by 
Eq.~(\ref{gi1}). As one can see in Fig.~2(a), 
some states which are detected by the
witnesses $\WW^{(i)}$ escape the detection via Eq.~(\ref{gi1}).

One can improve now the detection by using Criterion 4. Indeed, if 
we take an observable $M=\sum_i\mu_i\ketbra{BS_i}$ this criterion
requires for a separable state $\vr$:
\be
S^T_q(M)_\varrho \geq \frac{1-2^{1-q}}{q-1}.
\label{zqb}
\ee
This criterion, depending on $q$, can again be expressed in the 
coordinates $x,y,z.$ Note that for $q=2$ Eq.~(\ref{zqb}) is 
equivalent to Eq.~(\ref{gi1}). For two other values of $q$ 
Eq.~(\ref{zqb}) is plotted in Fig.~1(b) and 1(d). One can see that
the strength of the criterion increases with $q.$ This can also 
be proved analytically, and one can show that in the limit 
$q \rightarrow \infty$ Eq.~(\ref{zqb}) is equivalent to the witnesses
$\WW^{(i)}=1/2 \cdot \eins-\ketbra{BS_i}$ \cite{guehne2}. 

\section{Conclusion}
In conclusion, we have shown that separability conditions can be 
derived from variance based uncertainty relations 
as well as from entropic uncertainty relations. The investigation
of the resulting criteria showed that they are powerful tools 
for the detection of entanglement. 

We wish to thank Holger Hofmann, Philipp Hyllus, and Geza T\'oth 
for discussions.
This work has been supported by the DFG (Graduiertenkolleg 282 
and Schwerpunkt ``Quanten-Informationsverarbeitung'').



\bibliographystyle{aipproc}   


\end{document}